\documentclass[prd,twocolumn]{revtex4}
\usepackage{graphicx}
\usepackage{amssymb}
\usepackage{epstopdf}

\begin{document}

\title{Particle Dark Energy}
\author{Simon DeDeo}
\email{simon@astro.princeton.edu}
\affiliation{Department of Astrophysical Sciences, Princeton University, Princeton, New Jersey 08544, USA}

\begin{abstract}
We explore the physics of a gas of particles interacting with a condensate that spontaneously breaks Lorentz invariance. The equation of state of this gas varies from $1/3$ to less than $-1$ and can lead to the observed cosmic acceleration without requiring a vacuum energy. The particles are always stable. In our particular class of models these particles are fermions with a chiral coupling to the condensate. They may behave as relativistic matter at early times, produce a brief period where they dominate the expansion with $w<0$ today, and behave as matter at late time. There are no small parameters in our models, which generically lead to dark energy clustering and, depending on the choice of parameters, smoothing of small scale power.
\end{abstract}

\date{\today}

\maketitle

\section{Introduction}

A wide range of recent observations (see, \emph{e.g.}, Refs.~\cite{k03} and~\cite{s03}) suggest  a ``dark energy'' component that has an equation of state close to $-1$ and dominates the recent expansion history of the universe. The existence of the dark energy is a profound ``fine tuning'' problem for particle physics~\cite{w89}. Most solutions that can be expressed in terms of a Lagrangian invoke the presence of a non-zero vacuum energy~\cite{c01,c03}. 

Yet many physicists in the string theory community doubt it is possible to have any vacuum energy at all~\cite{w00}. Braneworld theories that invoke the presence of extra dimensions have other difficulties in producing a non-zero but small cosmological constant~\cite{cm01}.

We present an alternative mechanism for generating an accelerated expansion and negative equation of state that does not rely on a vacuum energy, and where no fine tuning is required. In our model, a fundamental field spontaneously breaks Lorentz invariance by developing a non-zero vector expectation value that thus ``picks'' a particular direction in spacetime. Just as in the case of the Higgs field in the electroweak theory the vacuum energy of the condensate is set to zero. 

The negative equation of state is produced by a gas of fermions that interacts with this Lorentz-violating condensate. At high energies, the fermion gas behaves like ordinary radiation; however, at temperatures comparable to the mass of the fermion, this gas behaves as if it has a negative equation of state.

We present the physics of particle dark energy in two parts. In Sec.~\ref{intro}, we derive the equation of motion for a fermion, $\psi$, coupled to a constant, time-like vector $b_{\mu}$, in curved spacetime. We then show how such a situation may arise in a fundamentally Lorentz invariant theory by the process of spontaneous symmetry breaking. We then show how a gas of such coupled $\psi$ particles may naturally lead to a negative equation of state, and possibly $w<-1$. 

In the remainder of the paper, Secs.~\ref{early} to~\ref{cluster}, we discuss realistic cosmological scenarios in which particle dark energy can explain the late-time acceleration.

\section{Lorentz-violating Couplings}
\label{intro}

We first consider the dynamics of a fermion field, $\psi$, coupled to a constant vector $b_{\mu}$ with dimensions of mass, a case that has been considered in flat spacetime by Ref.~\cite{ck97}. We take the Lagrangian to be 
\begin{equation}
\label{lag}
\mathcal{L} = \bar{\psi}\left(i\gamma^\mu D_\mu-m\right)\psi-\left(\bar{\psi}\gamma_5\gamma^\mu\psi\right)b_{\mu}.
\end{equation}Here $\psi$ is a fermion of mass $m$ (assumed to be small), and the operator $D_\mu$ represents the covariant derivative acting on the spinor.

We take the standard FRW metric to be
\begin{equation}
\label{metric}
ds^2=dt^2-a^2(t)d\vec{x}^2
\end{equation}
where $a(t)$ is the scale factor. We will take the vector $b_{\mu}$ to be time-like and, moreover, to be normal to the homogenous space-like hypersurfaces of the FRW time-slicing. Thus, we can write $b_{\mu}=(b,0,0,0)$, and the equation of motion for $\psi$ is
\begin{equation}
\label{eom}
\left(i\gamma_{\mu}\partial^{\mu}-\left[b\gamma_5-\frac{3i}{2}\frac{\dot{a}}{a}\right]\gamma^{0}-m\right)\psi=0,
\end{equation}
where $\dot{a}$ is $da/dt$ and the term proportional to the Hubble parameter, $\dot{a}/a$, comes from the spin connection term given by the metric in Eq.~\ref{metric} (see Ref.~\cite{bd87}.)  When $k\gg H$ we may discard terms proportional to $H$, thus recovering the flat-space equation of motion.

We write the wavefunction $\psi$ as a spinor multiplied by an exponential, being careful to allow for a non-standard dispersion relation between the time and spatial components of the momentum. Our ansatz is
\begin{eqnarray}
\label{motion}
\psi_\alpha(x,t) & = & u_\alpha(\omega,\vec{k})e^{-i[\omega_\alpha(\vec{k})t-a(t)\vec{k}\cdot\vec{x}]}, \\
\psi_\alpha(x,t) & = & v_\alpha(\omega,\vec{k})e^{i[\omega_\alpha(\vec{k})t-a(t)\vec{k}\cdot\vec{x}]},
\end{eqnarray}
where we have chosen to write $\vec{k}$ as the physical (not comoving) momentum, we have left the dependence of $\omega$ on $b$ implicit, and our ansatz is that $a(t)\vec{k}$ is constant.

After manipulation and again discarding terms proportional to $H$, one finds (taking the chiral basis, and up to a normalization),
\begin{equation}
\label{u}
u_+=\left( \begin{array}{c}-k+b+\omega_+(k) \\
0 \\
m \\
0
\end{array}\right);
u_-=\left( \begin{array}{c}0 \\
k+b+\omega_-(k) \\
0 \\
m
\end{array}\right),
\end{equation}
\begin{equation}
\label{v}
v_+=\left( \begin{array}{c}0 \\
-k+b-\omega_+(k) \\
0 \\
m
\end{array}\right);
v_-=\left( \begin{array}{c}k+b-\omega_-(k) \\
0 \\
m \\
0
\end{array}\right),
\end{equation}
\begin{equation}
\label{disp}
\omega_\pm=\sqrt{m^2+\left(k\mp b\right)^2},
\end{equation}
where we have chosen the direction of propagation to be along the $\hat{z}$-axis and the spinor subscript $\pm$ refers to states of positive and negative helicity.

The existence of these solutions implies that the physical three-momentum of a $\psi$ particle redshifts in the usual fashion. The particles do not ``decouple'' from the rest of the expansion and $k\propto a^{-1}$. This is to be expected from Noether's theorem, since spatial translations, and thus momentum, are preserved. As we shall see, conservation of stress-energy also requires this behavior.

Eq.~\ref{lag} describes a manifestly non-covariant theory because $b_\mu$ is fixed for all time. However, it is possible to arrive at such a Lagrangian as a low-energy approximation to a theory that develops a non-zero, vector-valued vacuum expectation value. This is a ``spontaneous'' breaking of local Lorentz invariance, analogous to the way in which the Higgs field spontaneously breaks the internal symmetry $SU(3)$. Early in the universe, at high temperatures, we expect both symmetries, Lorentz and $SU(3)$, to be restored.

To break Lorentz invariance spontaneously, one might imagine, for example, that $b_{\mu}$ is a vector field with a standard kinetic term, the above coupling to the field $\psi$, and a (gauge-violating) Mexican hat potential $V(b_{\mu}b^{\mu})$.

For the chiral coupling we have chosen in Eq.~\ref{lag} such a model would require, among other things, large dimensionless parameters. We will thus use instead the ``ghost matter'' model~\cite{ah04a,ah04b} as a concrete example of a Lorentz-violating condensate that can be written as a Lorentz-invariant fundamental theory. There are many recent models to which this analysis may be applied with potentially interesting results (\emph{e.g.}, Ref. \cite{a04}, which has spacelike condensates), and there is no fundamental reason why our model could not rely on vector- or tensor-valued condensates. Importantly, two recent works~\citep{klp03,blpr04} have discussed the possibility of spontaneous Lorentz violation via a scalar gradient, and the consequent breakdown of Lorentz invariance in the dispersion relation.

The Lagrangian for the ghost condensate, including our new fermion coupling, is
\begin{equation}
\label{lag2}
\begin{array}{l}
\mathcal{L} = P\left(X\right) + i\bar{\psi}\left(i\gamma^\mu D_\mu-m\right)\psi \medskip \\
-\frac{1}{F}\left(\bar{\psi}\gamma_5\gamma^\mu\psi\right)\partial_\mu\phi+\ldots.
\end{array}\end{equation}Here the field $\phi$ represents the scalar field condensate, and $\partial_\mu\phi\partial^\mu\phi$ is written as $X$ for brevity. The expectation of $\partial_\mu\phi$ will be our $b_\mu$. The term $F$ has dimensions of mass and is assumed to be of order the Planck scale.  The ellipsis represents higher order terms in the field $\phi$ which stabilize fluctuations in the $\phi$ field; their exact form will be unimportant for our investigation. 

The function $P$ is the non-standard kinetic term of the ghost matter model, and is chosen to have a minimum when the scalar field derivative, $\partial_\mu\phi$, is non-zero and time-like. 

Following Ref.~\cite{ah04a}, and by analogy with the Higgs mechanism, we take $P(X)$ to have a minimum at some non-zero, positive $X$. We expect this to be of the same order as the energy scale of the theory, $M$, and for simplicity we take $\partial_\mu\phi$ at the minimum to be $(M^2,0,0,0)$ in some frame. The most natural choice is to take this frame to be coincident with the CMB. Our parameter $b$ is then $M^2/F$.

As we shall see, a negative pressure will arise in this system even in the absence of a vacuum energy. In order to demonstrate this explicitly, we will take the value of $P(X)$ at the minimum to be zero. 

The presence of $\psi$-particles will perturb the field $\phi$. The classical equation of motion is
\begin{equation}
\label{cancel}
D_\mu\left(2FP^\prime(X)\partial^\mu\phi-\langle\bar{\psi}\gamma_5\gamma^{\mu}\psi\rangle\right)=0.
\end{equation}

We will consider situations in which there is an equal number density of $\psi$ particles with positive and with negative spin along any axis. In this case, the spatial components of the second term in Eq.~\ref{cancel} vanish, and no spatially varying excitations of the $\phi$ field are produced. (The creation of $\phi$ fluctuations by a collection of particles with net spin has been considered in the non-relativistic limit in Ref.~\cite{ah04c}.)

A simple solution to Eq.~\ref{cancel} can then be found: that for which the argument of $D_\mu$ vanishes, so that $2FP^\prime(X)\partial^\mu\phi$ is equal to $\langle\bar{\psi}\gamma_5\gamma^{\mu}\psi\rangle$. Conceptually, the $\phi$ field relaxes to its minimum as the $\psi$ particles dilute.

We can now calculate the stress-energy tensor of the full system by the Palatini procedure (see Sec. 8 of Ref.~\cite{bw57}, and Ref.~\cite{w01}.) We find
\begin{equation}
\begin{array}{l}
\label{t}
T^{\mu\nu}=-P(X)g^{\mu\nu}+\frac{i}{2}\left(\bar{\psi}\gamma^{(\mu}D^{\nu)}\psi-[D^{(\mu}\bar{\psi}]\gamma^{\nu)}\psi\right).
\end{array}
\end{equation}
Considering the second term, we see that the unusual dynamics of the $\phi$--$\psi$ coupling cancel each other, and the particle physics energy -- the energy conserved in interactions and that appears in the propagator -- and gravitational energy, $T^{00}$, of the $\psi$ particles coincide. Since this is seen only when one considers the complete system, including the contribution from the perturbed $\phi$ field, one might say that the system conspires to enforce the equivalence principle.

If $M$ is at least the electroweak scale, $F$ is the Planck scale and $m$ is of order $M^2/F$, then today the presence of $\psi$ particles changes the value of $\partial_0\phi$ by at most one part in $10^{60}$, and the first term $P(X)$ in Eq.~\ref{t} contributes an energy density $10^{60}$ times less than that of the $\psi$ particles themselves. This is not a ``fine tuning'': this extreme factor arises from the fact that the energy density of the universe today is much smaller than $M^4$. We may thus neglect this term in our later analysis.

\begin{figure}
\includegraphics[width=3.375in]{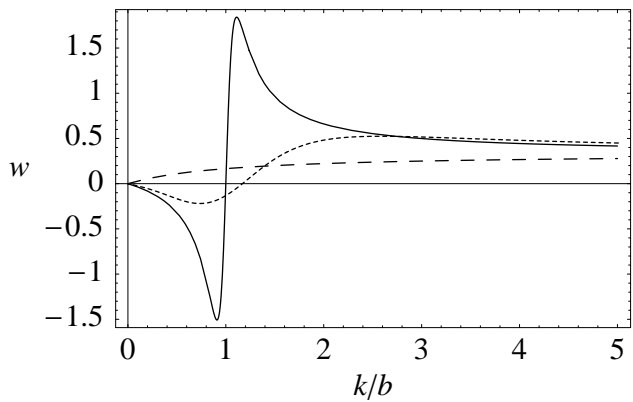}
\caption{The equation of state as a function of spatial momentum for a positive (solid line) and negative (dashed line) helicity gas with $\delta$-function distribution, and for a blackbody centered on $k$ for a positive helicity gas (dotted line), for the case $m=b/10$. The blackbody solution is a ``smoothed'' version of the $\delta$-function case.}
\label{w}
\end{figure}

We can now compute the equation of state for the $\psi$ particles as a function of their spatial momentum. We consider first a universe where all $\psi$-particles have the same momentum $k$, and are either all positive, or all negative, helicity. We find the equation of state for the two configurations to be

\begin{equation}
\label{eos}
w_\pm(k)=\frac{k(k\mp b)}{3(m^2+[k\mp b]^2)},
\end{equation}
which we plot in Fig.~\ref{w} for the case $m=b/10$.  We see that for positive helicity particles, the equation of state of the universe may become negative; indeed, $w$ may even become less than negative one, something that can not be achieved with ordinary scalar field condensates.

It is important to note that the negative equation of state does not come from a vacuum energy; the potential $P(X)$, as noted above, contributes a negligible amount to the energy density of the universe. Instead, the negative pressure arises in the gas of particles itself. One way to see this is as the conservation of spatial momentum requiring that the $\psi$ particles redshift, combined with the fact that for $k<b$, the energy per particle increases with decreasing $k$. The requirement that the $\psi$ particles redshift can be found by requiring that $T^{\mu\nu}$ [Eq.~\ref{t}] be conserved.

The group velocity, $\vec{v}_g$ is
\begin{equation}
\label{vg}
v_{g,\pm}=\frac{d\omega}{dk}=\frac{k\mp b}{\sqrt{m^2+(k\mp b)^2}},
\end{equation}
and for positive helicity particles, when $k$ is less than $b$, $\vec{v}_g$ and the spatial momentum, $\vec{k}$, point in opposite directions. Thus another, intuitive way to understand the emergence of negative pressure is to note that a particle leaving a volume in one direction carries carries with it a momentum pointed in the opposite direction. 

A somewhat analogous situation to the one in this paper can arise in MHD systems, where a magnetosonic wave can develop a negative pressure~\cite{d70}. Note that the behavior of the group velocity in our model is qualitatively different from what happens near an absorbtion resonance~\cite{mv04c}. 

Our discovery of a negative $w$ is related to a previous result that found a changing, though not a negative, $w$ from an analysis of the dispersion relation for a gas of particles in a cosmological scenario involving Lorentz violation~\citep{jm01b}. 

Our model, relying as it does on a particular mechanism for spontaneous Lorentz violation, is relevant, though not identical, to a wide class of mechanisms developed by numerous other authors. For example,~\cite{jm01a} examined a Lorentz-violating system where the vector, $u^{\mu}$, was a fundamental field of fixed norm and not, as in our case, the spatial gradient of a scalar field. 

In that situation, important cosmological bounds exist from the coupling of $u^{\mu}$ to gravity; in particular, the possibility of an effectively changing Newton's constant $G$ arises; see, \emph{e.g.}, \cite{ejm04} and references therein. Our model's (minimal) coupling to gravity, which we have derived above, is more complicated.

Later papers that consider a fixed-norm $u^{\mu}$, including \cite{jm04}, have further developed the theory, considering the nature of perturbations in the $u$ field. In our model, as discussed earlier in the section, the high mass scale of the $\phi$ field means that late-time production of perturbations will be negligible; these perturbations certaintly exist, however, and their nature is left for future investigation.

\begin{figure}
\includegraphics[width=3.375in]{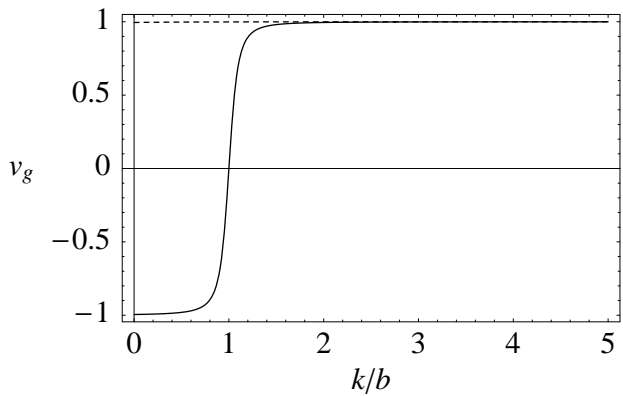}
\caption{The group velocity, $v_g$, for positive helicity (solid line) and negative helicity (dotted line) particles, for the case $m=b/10$.}
\label{group}
\end{figure}

In Fig.~\ref{group} we plot $v_g$ as a function of $k$. As the system is non-dissipative, the sound speed, $c_s$, is just $v_g/\sqrt{3}$. Comparing this with Fig.~\ref{w}, we see that $c_s\rightarrow0$ as $w\rightarrow0$; when the $\psi$ particles have the same equation of state as ordinary dark matter they also have the same clustering properties in the linear regime. This suggests that it may be possible for the $\psi$ particles to serve some of the roles of both dark energy and dark matter, a possibility we examine below in Sec.~\ref{late}.

Generically, we expect the $\psi$ particles to be in thermal equilibrium at some very early time. At some later time, the $\psi$ particles decouple, and their distribution then redshifts. The decoupling could happen either early, \emph{i.e.}, when the $\psi$ particles have $T\gg m$, or late, when $T\approx m$. We consider the two cases separately, now in the context of a realistic cosmology.

\section{Early decoupling} 
\label{early}

Let us first consider the case where the $\psi$ particles decouple while still relativistic. In this scenario, a Fermi-Dirac distribution simply redshifts without changing its shape.

From an examination of Eq.~\ref{eos}, we see that for the equation of state to become negative, $b$ must  larger than $m$. This restricts the number of e-folds in $k$ over which the particles may be distributed; as $b$ becomes larger, $w$ can be more negative, but over a smaller $\Delta k/k$.

The width in $k$ over which $w$ is significantly negative is $k_{\rm finish}/k_{\rm start}\approx(b+m)/(b-m)\lesssim1.2$. Since the Fermi-Dirac distribution at large temperatures is significantly wider than this, the possibility of a very negative $w$, and universal acceleration, is hidden. This is shown in  Fig.~\ref{w}, where the equation of state as a function of scale factor for a redshifting Fermi-Dirac distribution of positive helicity particles is plotted as the dotted line.

We find a brief period of mild dark energy behavior between the radiation and matter-like eras. Including both positive and negative helicity states reduces the positive and negative excursions of the equation of state further. 

It may be checked that if the $\psi$ particles are subdominant to cold dark matter at small scale factors, when they have $w\approx1/3$, they are always subdominant. Thus there is no way to produce a significant period of matter domination before the onset of a negative equation of state. Early decoupling, then, is not a promising model for the late-time acceleration, although it may have some applicability in the early universe.

\section{Late decoupling}
\label{late}

We now consider the case opposite to early decoupling. We assume that, due to a significant self-interaction term, the $\psi$ particles decouple only after they become non-relativistic. We approximate this situation by assuming that before the interval between interactions is equal to $1/H$, the particles remain in exact thermal equilibrium, at a temperature $T$ corresponding to the energy density of particles. When the interaction rate is equal to $1/H$, we assume that the particles gradually drop out of equilibrium over a finite range of scale factor.

We find that before decoupling the positive helicity particles form a narrowing clump in the ``well'' of the dispersion relation at $k=b$ where $w=0$. The population of negative helicity particles is strongly suppressed. 

While the $\psi$ particles remain in thermal equilibrium the universe appears to be undergoing a standard transition from radiation to matter domination. When the particles drop out of equilibrium, the very narrow distribution at $k=b$ redshifts, and we recover a strongly negative equation of state for roughly one e-fold. At very large $a$, the universe is matter-dominated.

\begin{figure}
\includegraphics[width=3.375in]{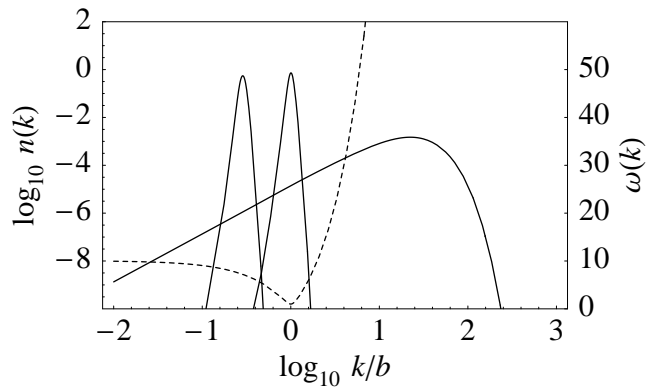}
\caption{A series of ``snapshots'' of the particle distribution, $n(k)$  (solid lines) for the case $m=1$, $b=10$. The dashed line is the dispersion relation, $\omega(k)$. Reading the solid lines chronologically from right to left, this figure shows how a high-temperature Fermi-Dirac distribution cools, and particles clump in the well of the dispersion relation at $k=b$ and then redshift after decoupling.}
\label{snap}
\end{figure}

Fig.~\ref{snap} illustrates this with three overlaid ``snapshots'' of the $\psi$ particle distribution. The figure is to be read right to left in chronological order. The first snapshot is well before decoupling when $T\approx100m$ and the particles behave approximately like radiation. This is the same as the early decoupling case, since relativistic particles maintain a thermal profile even after decoupling. This distribution is nearly indistinguishable from that of photons in thermal equilibrium at the same temperature.

The second snapshot is before decoupling but after the universe has cooled further; here $T\approx0.25m$ and $w\approx0$. The evolution has diverged from the early decoupling case, and the particle distribution forms a narrowing clump at the bottom of the well.

The third snapshot is after the system has decoupled. The distribution has redshifted towards $k=0$, and now $w<0$. We take decoupling to happen over a finite range of scale factors: different particles leave the bottom of the well at slightly different times, so the distribution is slightly wider than just before decoupling.

We can thus have an arbitrarily long period of matter domination between the radiation and dark energy dominated eras. As before, the dark energy era lasts only a brief period of time, and now appears soon after the interaction time becomes of order $1/H$. In contrast to the early decoupling case above, however, a previously subdominant $\psi$-fluid component can grow to become the dominant contribution to the stress-energy.

Let us now consider the plausibility of a late decoupling around $z\approx0$, to explain the contemporary acceleration. We require that
\begin{equation}
\label{equi}
n_0 \sigma v \sim H_0
\end{equation}
where $n_0$ is the number density of particles, of order $\rho_0/m$, $\sigma$ is the cross section, which we take equal to $1/m^2$, and $v$ is the average particle speed.

It remains to determine $v$. To a first approximation, we can take $v$ approximately equal to $c$ at all times. This seems counterintuitive, because if a particle is kicked into a state $k\approx b$, its group velocity is very small. The particle may drop out of equilibrium and start to redshift. 

However, the unusual nature of the group velocity implies that a redshifting particle at $k=b$ finds its velocity increasing. The steepness of the group velocity curve around this point means that a particle need only redshift by a factor of $1+0.1m/b$ in order for its velocity to rise from zero to $\approx0.1c$. After redshifting this small amount, it can start to reestablish the equilibrium suggested by setting $v$ to unity in Eq.~\ref{equi} above.

Setting $v$ above to unity, then, we can find the maximum mass of the $\psi$ particle such that decoupling happens around $t_0$ today. Nucleosynthesis gives us the constraint that, if the $\psi$ particles are to make up a non-negligible portion of the energy density today, they must enter the $w=0$ phase before $T\approx1$ MeV. We find, to an order of magnitude, $m\lesssim100$ MeV if the particles are to decouple today, which easily satisfies this bound. Setting the fraction of $\psi$ particles to $\Omega_\psi\neq1$, or changing the cross-section by a factor $\alpha^2$ lowers this mass estimate only by a factor $\alpha^{2/3}\Omega_\psi^{1/3}$.

\begin{figure}
\includegraphics[width=3.375in]{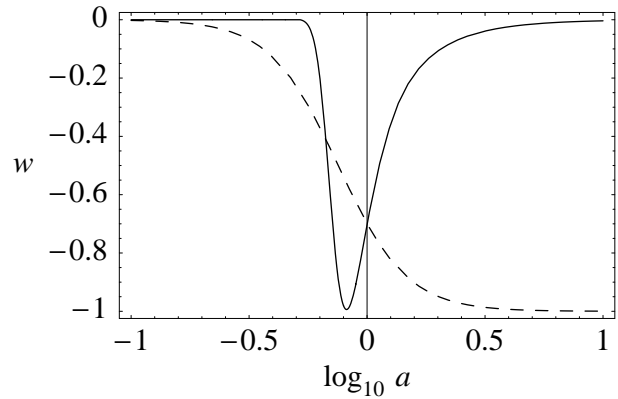}
\caption{The equation of state of the entire universe (\emph{i.e.}, including all components) as a function of scale factor in a $\Psi$CDM scenario (solid line.) Here we have taken the decoupling to happen between $0.20\lesssim z\lesssim0.62$, we have taken the $\Omega_{\psi,i}=0.08$ at $z>1$, and $b=M^2/F=400m$. The $d_{\rm LSS}$, $h$ and $q_0$ for this model agree with that of a fiducial $\Lambda$CDM model, plotted for reference as a dashed line.}
\label{decouple}
\end{figure}

We thus consider a ``$\Psi$CDM'' model, a flat universe with similar energy densities of $\psi$ matter and some cold dark matter particle. Approximating the $\psi$ decoupling as happening over some finite range in scale factor $a_{\rm dec}$, we plot in Fig.~\ref{decouple} the equation of state of the universe for $\Psi$CDM. with $\Lambda$CDM plotted for comparison. Following the discussion above, we take decoupling to happen over a range of redshifts; the particular choice here is $0.20\lesssim z\lesssim0.62$, and we take $\Omega_{\psi}$ before decoupling to be $0.08$.

We have chosen parameters so that both models have the same $H_0$, $d_{\rm lss}$ and deceleration parameter $q_0$; there are many different combination of decoupling redshift, $\Omega_{\psi}$ and $b$ that can satisfy this requirement. Making the decoupling process take longer widens the ``well'' of the effective equation of state in Fig.~\ref{decouple}. 

The simple analysis above suggests that it is possible to describe the late-time acceleration of the universe with a very simple model that relies on a late-time decoupling of the $\psi$ particles. There is no fine-tuning required: the $\psi$ particles must be subdominant to satisfy large scale structure constraints, but their contribution to the density before decoupling may be between $1\%$ and $10\%$ of the CDM. If our $\psi$ particles are to have mass near $100$ MeV, then if $F$ is the Planck scale, the intermediate scale $M$ is $\sim10^{9}$ GeV.

The fraction of energy density contained in $\psi$ particles at high redshift is non-negligible, and depending on choice of parameters can be of order tens of percent. At high redshifts, the $\psi$ particles act like ordinary dark matter, and can cluster. Closer to today, during the contemporary acceleration, they have a near-unity group velocity. In many ways, then, the phenomenology of our model is similar to a decaying dark matter scenario~\cite{cen01}, where some fraction of the dark matter turns into ultra-relativistic particles, smoothing out small scale power. 

It is this freestreaming that makes it difficult for the $\psi$ particles to be both dark energy and all of the dark matter, although more sophisticated models, with a more detailed treatment of the unusual decoupling process and the evolution of perturbations, may allow us to unify these two mysterious components. We consider this an open question; we discuss the clustering properties of this model in greater detail in Sec.~\ref{cluster} below. 

\section{Decaying Dark Matter}

We believe that the late-time decoupling scenario is the most promising particle dark energy model of the kinds discussed in this paper. However, it is instructive to consider other scenarios, in part so as to understand the particle physics aspects of the $\psi$ fluid.

Our second toy model is for the $\psi$ particles to be produced during the late-time decay of a separate dark matter particle. Our so-called ``D$\Psi$CDM'' model is a modified version of the decaying CDM models first proposed in Ref.~\cite{cen01}, where in D$\Psi$CDM the decay products are $\psi$ particles.

Let us assume that the dark matter particles are described by a single scalar field $\chi$ of mass $m_\chi$ that may decay via a $\psi$-number conserving interaction with an exponential decay time $\tau$ we take to be of order $1/H_0$. We assume the $\chi$ particles are cold.

If the mass of the $\chi$ particle, $m_\chi$, is greater than $\sqrt{m^2+b^2}$, the $\chi$ particle decay products will include both positive and negative helicity $\psi$ states. However, if $m_\chi$ is less than $\sqrt{m^2+b^2}$, decay into negative helicity states is energetically forbidden.

Let us consider this case and take, for simplicity, $m_\chi$ equal to $m$, the mass of the $\psi$ particle. Then, the $\chi$ decay products end up in a narrow distribution of positive helicity $\psi$ particle states at $k=b$. As the $\psi$ particles are produced, they redshift to $k<b$ and, as above, develop negative pressure.

\begin{figure}
\includegraphics[width=3.375in]{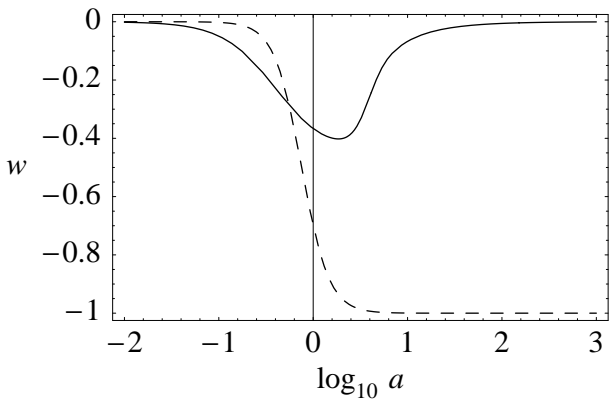}
\caption{The equation of state of the entire universe as a function of scale factor in a D$\Psi$CDM scenario (solid line.) Here we have taken $\tau=2t_0$ and $b=45m$. The $d_{\rm LSS}$ and $h$ for both models agree with that of the $\Lambda$CDM model, plotted for reference as a dashed line.}
\label{decay}
\end{figure}

We approximate the universe as beginning with only $\chi$ particles; the $\psi$ population is built up as the $\chi$ particles decay with e-folding time $\tau$. Fig.~\ref{decay} shows the equation of state of the universe, as a function of redshift, for a model where we have taken $\tau$ to be of order the age of the universe, $t_0$, and have chosen the mass scale, $b$, such that the distance to last scattering, and the Hubble constant today, are the same as a fiducial $\Lambda$CDM, which we plot for comparison.

Though both models, D$\Psi$CDM and $\Lambda$CDM, in Fig.~\ref{decay} have the same $d_{\rm LSS}$ and $h$, they give different values for the deceleration parameter, $q_0$. For D$\Psi$CDM, $q_0\approx-0.05$, while for $\Lambda$CDM $q_0\approx-0.55$.

We find thus that this decaying dark matter model naturally produces a period of accelerated expansion in the low-redshift universe even if only a small fraction of dark matter has decayed. This is because the equation of state for the $\psi$ particle population is very negative just around $k\approx b$; the $\psi$ particles grow rapidly in comoving energy as soon as they are produced.

The position of the turnover depends both on the relative values of $b$ and $m$, and on the dark matter lifetime. One can keep the position of the turnover fixed by increasing both $b/m$ and the dark matter lifetime, which the overall shape remains unchanged. The equation of state returns to zero after most of the dark matter has decayed and the majority of $\psi$ particles have redshifted to $k\ll b$.

There are three mass scales in our theory: $M$, $F$ and $m$, where $m$ is supposed to be near $M^2/F$. If we take $F$ to be the Planck scale, then if $M$ is the GUT scale, the dark matter mass is greater than $10^{12}$ GeV. Alternatively, if our dark matter is an Axion condensate with $m\sim 10^{-5}$~eV, $M$ may be as low as $100$ TeV, roughly the electroweak scale.

\section{Evolution of Perturbations}
\label{cluster}

The evolution of perturbations in particle dark energy models remains to be studied; however, broad statements may be made. Both $\Psi$CDM and D$\Psi$CDM make definite predicitions about the growth of large scale structure. 

First, both predict the signatures of dark energy clustering. In the $\Psi$CDM model, the sound speed, related to $v_g$, the group velocity, goes to zero with $w$ just before the acceleration, so that the $\psi$ fluid perturbations may grow like ordinary matter before their equation of state becomes negative. In D$\Psi$CDM, the progenitors of the $\psi$ particles may cluster like ordinary dark matter. The reader is cautioned that the behavior of clustering dark energy in the non-linear regime requires a separate treatment~\cite{mv04b}.

Whether and how dark energy clusters remains an urgent question both for observation and theory~\cite{bd04,mv04}; tantalizingly, we may have seen evidence for it in the deficit of low-$\ell$ power in the CMB~\cite{dcs03}.

Second, both models may have direct relevance to the possible excess small scale structure problem (see, \emph{e.g.}, Ref.~\cite{dssw01}.) In $\Psi$CDM, the $\psi$ particles may make up a substantial fraction (of order tens of percent) of the energy density of the universe. When they decouple, their group velocity rises rapidly to unity, and they may stream out of overdensities on small scales, smoothing out small scale power.

This is just as in the first decaying dark matter models~\cite{cen01}. D$\Psi$CDM, of course, actually is a model with decay: there dark matter turns into ultra-relativistic particles that then develop a negative equation of state.

\section{Conclusions}
\label{conc}

We have discussed the phenomenon of Lorentz-violating couplings in the context of an FRW curved spacetime, and have demonstrated how a fermion coupled to a spontaneous Lorentz-violating condensate can produce a strongly negative equation of state. We have also presented two models, one requiring a late-time decoupling (``$\Psi$CDM''), and the other requiring a late-time decay (``D$\Psi$CDM''), that, at least in outline form, can explain the contemporary acceleration. In particular, the $\Psi$CDM model requires no fine-tuning, and involves natural cross-sections.

The rich set of behaviors, and the number of scenarios that can be constructed even from such a toy model as Eq.~\ref{lag2} suggests that the question of using different Lorentz breaking mechanisms, such as a non-zero expectation value for a vector or tensor field, deserves further investigation. There are also a wide range of possible couplings between such a Lorentz-violating expectation value and various kinds of matter, including not only fermion couplings, but also derivative couplings to scalar field particles, and direct couplings to particles of higher spin. 

A catalogue of Lorentz violating couplings to the standard model has been investigated in Ref.~\cite{ck98}; here, since we are coupling to the dark sector, we may use much larger couplings without violating Lorentz-violation bounds from atomic or accelerator physics. A first indicator of whether or not a system will generate a negative equation of state in any model of interest will be a turn-over in the particle's group velocity, Eq.~\ref{vg}.

Lorentz violation has long been a subject of study in the particle and atomic physics community; we are only just beginning to discover its cosmological implications~\cite{bc00}. Many recent astrophysical studies have focused on high energy signatures, such as the GZK cutoff~\cite{o00}, and in the context of loop quantum gravity or non-commutative geometry~\cite{amu02,bm03,sm01}. Here, in contrast, we have demonstrated how Lorentz-violating couplings can lead to negative pressures in the context of a spontaneous breaking of Lorentz invariance, and we have shown how such violation may have astrophysical implications at low energies.

Spontaneous Lorentz violation has long been considered as a feature of string theory for the bosonic string and the superstring~\cite{ks89a,ks89b,b97}. It may also be a generic aspect of braneworld models, where Lorentz violation occurs in the bulk and may be seen on the brane as well~\cite{f03}. Finally, the role of the universal time-like vector could even be played by the same mechanism underlying the thermodynamic ``arrow of time.''

Because of the generic nature of our discussion, we expect the essence of our conclusions to hold true for these more elaborate scenarios when a time-like Lorentz-violating condensate couples to the dark sector. As this paper has shown, there is a rich set of phenomena to be discovered that may potentially resolve long-standing theoretical and observational questions.
\\
\section{Acknowledgments}

I thank D. N. Spergel for support throughout this project, and P. J. Steinhardt, H. Verlinde and my anonymous referee for very helpful conversations. I acknowledge the support of NASA Theory Award NNG04GK55G, ``Implications of CMB Observations.''


\begin{thebibliography}{39}
\expandafter\ifx\csname natexlab\endcsname\relax\def\natexlab#1{#1}\fi
\expandafter\ifx\csname bibnamefont\endcsname\relax
  \def\bibnamefont#1{#1}\fi
\expandafter\ifx\csname bibfnamefont\endcsname\relax
  \def\bibfnamefont#1{#1}\fi
\expandafter\ifx\csname citenamefont\endcsname\relax
  \def\citenamefont#1{#1}\fi
\expandafter\ifx\csname url\endcsname\relax
  \def\url#1{\texttt{#1}}\fi
\expandafter\ifx\csname urlprefix\endcsname\relax\def\urlprefix{URL }\fi
\providecommand{\bibinfo}[2]{#2}
\providecommand{\eprint}[2][]{\url{#2}}

\bibitem[{\citenamefont{{Knop \emph{et al.}}}(2003)}]{k03}
\bibinfo{author}{\bibfnamefont{R.~A.} \bibnamefont{{Knop \emph{et al.}}}},
  \bibinfo{journal}{ApJ} \textbf{\bibinfo{volume}{598}}, \bibinfo{pages}{102}
  (\bibinfo{year}{2003}).

\bibitem[{\citenamefont{{Spergel \emph{et al.}}}(2003)}]{s03}
\bibinfo{author}{\bibfnamefont{D.~N.} \bibnamefont{{Spergel \emph{et al.}}}},
  \bibinfo{journal}{ApJS} \textbf{\bibinfo{volume}{148}}, \bibinfo{pages}{175}
  (\bibinfo{year}{2003}).

\bibitem[{\citenamefont{{Weinberg}}(1989)}]{w89}
\bibinfo{author}{\bibfnamefont{S.}~\bibnamefont{{Weinberg}}},
  \bibinfo{journal}{Rev. Mod. Phys.} \textbf{\bibinfo{volume}{61}},
  \bibinfo{pages}{1} (\bibinfo{year}{1989}).

\bibitem[{\citenamefont{{Carroll}}(2001)}]{c01}
\bibinfo{author}{\bibfnamefont{S.}~\bibnamefont{{Carroll}}},
  \bibinfo{journal}{Living Rev. Rel.} \textbf{\bibinfo{volume}{4}}
  (\bibinfo{year}{2001}), \bibinfo{note}{cited on 21 May 2004}.

\bibitem[{\citenamefont{{Carroll}}(2003)}]{c03}
\bibinfo{author}{\bibfnamefont{S.}~\bibnamefont{{Carroll}}}
  (\bibinfo{year}{2003}), \eprint{astro-ph/0310342}.

\bibitem[{\citenamefont{{Witten}}(2000)}]{w00}
\bibinfo{author}{\bibfnamefont{E.}~\bibnamefont{{Witten}}}
  (\bibinfo{year}{2000}), \eprint{hep-ph/0002297}.

\bibitem[{\citenamefont{{Carroll} and {Mersini}}(2001)}]{cm01}
\bibinfo{author}{\bibfnamefont{S.}~\bibnamefont{{Carroll}}} \bibnamefont{and}
  \bibinfo{author}{\bibfnamefont{L.}~\bibnamefont{{Mersini}}},
  \bibinfo{journal}{\prd} \textbf{\bibinfo{volume}{64}},
  \bibinfo{pages}{124008} (\bibinfo{year}{2001}).

\bibitem[{\citenamefont{{Colladay} and {{Kosteleck\'{y}}, V. A.}}(1997)}]{ck97}
\bibinfo{author}{\bibfnamefont{D.}~\bibnamefont{{Colladay}}} \bibnamefont{and}
  \bibinfo{author}{\bibnamefont{{{Kosteleck\'{y}}, V. A.}}},
  \bibinfo{journal}{Phys. Rev. D} \textbf{\bibinfo{volume}{55}},
  \bibinfo{pages}{6760} (\bibinfo{year}{1997}).

\bibitem[{\citenamefont{{Barut} and {Duru}}(1987)}]{bd87}
\bibinfo{author}{\bibfnamefont{A.~O.} \bibnamefont{{Barut}}} \bibnamefont{and}
  \bibinfo{author}{\bibfnamefont{I.~H.} \bibnamefont{{Duru}}},
  \bibinfo{journal}{\prd} \textbf{\bibinfo{volume}{36}}, \bibinfo{pages}{3705}
  (\bibinfo{year}{1987}).

\bibitem[{\citenamefont{{Arkani-Hamed}
  et~al.}(2003)\citenamefont{{Arkani-Hamed}, {Hsin-Chia}, and
  {Mukohyama}}}]{ah04a}
\bibinfo{author}{\bibfnamefont{N.}~\bibnamefont{{Arkani-Hamed}}},
  \bibinfo{author}{\bibfnamefont{C.}~\bibnamefont{{Hsin-Chia}}},
  \bibnamefont{and}
  \bibinfo{author}{\bibfnamefont{S.}~\bibnamefont{{Mukohyama}}}
  (\bibinfo{year}{2003}), \eprint{hep-th/0312099}.

\bibitem[{\citenamefont{{Arkani-Hamed}
  et~al.}(2004{\natexlab{a}})\citenamefont{{Arkani-Hamed}, {Creminelli},
  {Mukohyama}, and {Zaldarriaga}}}]{ah04b}
\bibinfo{author}{\bibfnamefont{N.}~\bibnamefont{{Arkani-Hamed}}},
  \bibinfo{author}{\bibfnamefont{P.}~\bibnamefont{{Creminelli}}},
  \bibinfo{author}{\bibfnamefont{S.}~\bibnamefont{{Mukohyama}}},
  \bibnamefont{and}
  \bibinfo{author}{\bibfnamefont{M.}~\bibnamefont{{Zaldarriaga}}},
  \bibinfo{journal}{JCAP} \textbf{\bibinfo{volume}{4}}, \bibinfo{pages}{1}
  (\bibinfo{year}{2004}{\natexlab{a}}).

\bibitem[{\citenamefont{{Armend\'{a}riz-Pic\'on}}(2004)}]{a04}
\bibinfo{author}{\bibfnamefont{C.}~\bibnamefont{{Armend\'{a}riz-Pic\'on}}}
  (\bibinfo{year}{2004}), \eprint{astro-ph/0405267}.

\bibitem[{\citenamefont{{Kosteleck{\' y}}
  et~al.}(2003)\citenamefont{{Kosteleck{\' y}}, {Lehnert}, and
  {Perry}}}]{klp03}
\bibinfo{author}{\bibfnamefont{V.~A.} \bibnamefont{{Kosteleck{\' y}}}},
  \bibinfo{author}{\bibfnamefont{R.}~\bibnamefont{{Lehnert}}},
  \bibnamefont{and} \bibinfo{author}{\bibfnamefont{M.~J.}
  \bibnamefont{{Perry}}}, \bibinfo{journal}{\prd}
  \textbf{\bibinfo{volume}{68}}, \bibinfo{pages}{123511}
  (\bibinfo{year}{2003}).

\bibitem[{\citenamefont{{Bertolami} et~al.}(2004)\citenamefont{{Bertolami},
  {Lehnert}, {Potting}, and {Ribeiro}}}]{blpr04}
\bibinfo{author}{\bibfnamefont{O.}~\bibnamefont{{Bertolami}}},
  \bibinfo{author}{\bibfnamefont{R.}~\bibnamefont{{Lehnert}}},
  \bibinfo{author}{\bibfnamefont{R.}~\bibnamefont{{Potting}}},
  \bibnamefont{and}
  \bibinfo{author}{\bibfnamefont{A.}~\bibnamefont{{Ribeiro}}},
  \bibinfo{journal}{\prd} \textbf{\bibinfo{volume}{69}},
  \bibinfo{pages}{083513} (\bibinfo{year}{2004}).

\bibitem[{\citenamefont{{Arkani-Hamed}
  et~al.}(2004{\natexlab{b}})\citenamefont{{Arkani-Hamed}, {Cheng}, {Luty}, and
  {Thaler}}}]{ah04c}
\bibinfo{author}{\bibfnamefont{N.}~\bibnamefont{{Arkani-Hamed}}},
  \bibinfo{author}{\bibfnamefont{H.}~\bibnamefont{{Cheng}}},
  \bibinfo{author}{\bibfnamefont{M.}~\bibnamefont{{Luty}}}, \bibnamefont{and}
  \bibinfo{author}{\bibfnamefont{J.}~\bibnamefont{{Thaler}}}
  (\bibinfo{year}{2004}{\natexlab{b}}), \eprint{hep-ph/0407034}.

\bibitem[{\citenamefont{{Brill} and {Wheeler}}(1957)}]{bw57}
\bibinfo{author}{\bibfnamefont{D.~R.} \bibnamefont{{Brill}}} \bibnamefont{and}
  \bibinfo{author}{\bibfnamefont{J.~A.} \bibnamefont{{Wheeler}}},
  \bibinfo{journal}{Rev. Mod. Phys.} \textbf{\bibinfo{volume}{29}},
  \bibinfo{pages}{465} (\bibinfo{year}{1957}).

\bibitem[{\citenamefont{{Weldon}}(2001)}]{w01}
\bibinfo{author}{\bibfnamefont{H.~A.} \bibnamefont{{Weldon}}},
  \bibinfo{journal}{\prd} \textbf{\bibinfo{volume}{63}},
  \bibinfo{pages}{104010} (\bibinfo{year}{2001}).

\bibitem[{\citenamefont{{Dewar}}(1970)}]{d70}
\bibinfo{author}{\bibfnamefont{R.~L.} \bibnamefont{{Dewar}}},
  \bibinfo{journal}{Phys. Fluids} \textbf{\bibinfo{volume}{13}},
  \bibinfo{pages}{2710} (\bibinfo{year}{1970}).

\bibitem[{\citenamefont{{de la Macorra} and {Vucetich}}(2004)}]{mv04c}
\bibinfo{author}{\bibfnamefont{A.}~\bibnamefont{{de la Macorra}}}
  \bibnamefont{and}
  \bibinfo{author}{\bibfnamefont{H.}~\bibnamefont{{Vucetich}}},
  \bibinfo{journal}{Journal of Cosmology and Astro-Particle Physics}
  \textbf{\bibinfo{volume}{9}}, \bibinfo{pages}{12} (\bibinfo{year}{2004}).

\bibitem[{\citenamefont{{Jacobson} and
  {Mattingly}}(2001{\natexlab{a}})}]{jm01b}
\bibinfo{author}{\bibfnamefont{T.}~\bibnamefont{{Jacobson}}} \bibnamefont{and}
  \bibinfo{author}{\bibfnamefont{D.}~\bibnamefont{{Mattingly}}},
  \bibinfo{journal}{Phys. Rev. D} \textbf{\bibinfo{volume}{63}},
  \bibinfo{pages}{041502} (\bibinfo{year}{2001}{\natexlab{a}}),
  \bibinfo{note}{(R)}.

\bibitem[{\citenamefont{{Jacobson} and
  {Mattingly}}(2001{\natexlab{b}})}]{jm01a}
\bibinfo{author}{\bibfnamefont{T.}~\bibnamefont{{Jacobson}}} \bibnamefont{and}
  \bibinfo{author}{\bibfnamefont{D.}~\bibnamefont{{Mattingly}}},
  \bibinfo{journal}{Phys. Rev. D} \textbf{\bibinfo{volume}{64}},
  \bibinfo{pages}{024028} (\bibinfo{year}{2001}{\natexlab{b}}).

\bibitem[{\citenamefont{Eling et~al.}(2004)\citenamefont{Eling, Jacobson, and
  Mattingly}}]{ejm04}
\bibinfo{author}{\bibfnamefont{C.}~\bibnamefont{Eling}},
  \bibinfo{author}{\bibfnamefont{T.}~\bibnamefont{Jacobson}}, \bibnamefont{and}
  \bibinfo{author}{\bibfnamefont{D.}~\bibnamefont{Mattingly}}
  (\bibinfo{year}{2004}), \eprint{gr-qc/0410001}.

\bibitem[{\citenamefont{Jacobson and Mattingly}(2004)}]{jm04}
\bibinfo{author}{\bibfnamefont{T.}~\bibnamefont{Jacobson}} \bibnamefont{and}
  \bibinfo{author}{\bibfnamefont{D.}~\bibnamefont{Mattingly}},
  \bibinfo{journal}{Phys. Rev. D} \textbf{\bibinfo{volume}{70}},
  \bibinfo{pages}{024003} (\bibinfo{year}{2004}).

\bibitem[{\citenamefont{{Cen}}(2001)}]{cen01}
\bibinfo{author}{\bibfnamefont{R.}~\bibnamefont{{Cen}}},
  \bibinfo{journal}{ApJL} \textbf{\bibinfo{volume}{546}}, \bibinfo{pages}{L77}
  (\bibinfo{year}{2001}).

\bibitem[{\citenamefont{{Nunes} and {Mota}}(2004)}]{mv04b}
\bibinfo{author}{\bibfnamefont{N.~J.} \bibnamefont{{Nunes}}} \bibnamefont{and}
  \bibinfo{author}{\bibfnamefont{D.~F.} \bibnamefont{{Mota}}}
  (\bibinfo{year}{2004}), \eprint{astro-ph/0409481}.

\bibitem[{\citenamefont{{Bean} and {Dor{\' e}}}(2004)}]{bd04}
\bibinfo{author}{\bibfnamefont{R.}~\bibnamefont{{Bean}}} \bibnamefont{and}
  \bibinfo{author}{\bibfnamefont{O.}~\bibnamefont{{Dor{\' e}}}},
  \bibinfo{journal}{\prd} \textbf{\bibinfo{volume}{69}},
  \bibinfo{pages}{083503} (\bibinfo{year}{2004}).

\bibitem[{\citenamefont{{Mota} and {van de Bruck}}(2004)}]{mv04}
\bibinfo{author}{\bibfnamefont{D.~F.} \bibnamefont{{Mota}}} \bibnamefont{and}
  \bibinfo{author}{\bibfnamefont{C.}~\bibnamefont{{van de Bruck}}},
  \bibinfo{journal}{A\&A} \textbf{\bibinfo{volume}{421}}, \bibinfo{pages}{71}
  (\bibinfo{year}{2004}).

\bibitem[{\citenamefont{{DeDeo} et~al.}(2003)\citenamefont{{DeDeo}, {Caldwell},
  and {Steinhardt}}}]{dcs03}
\bibinfo{author}{\bibfnamefont{S.}~\bibnamefont{{DeDeo}}},
  \bibinfo{author}{\bibfnamefont{R.~R.} \bibnamefont{{Caldwell}}},
  \bibnamefont{and} \bibinfo{author}{\bibfnamefont{P.~J.}
  \bibnamefont{{Steinhardt}}}, \bibinfo{journal}{\prd}
  \textbf{\bibinfo{volume}{67}}, \bibinfo{pages}{103509}
  (\bibinfo{year}{2003}).

\bibitem[{\citenamefont{{Dav{\' e}} et~al.}(2001)\citenamefont{{Dav{\' e}},
  {Spergel}, {Steinhardt}, and {Wandelt}}}]{dssw01}
\bibinfo{author}{\bibfnamefont{R.}~\bibnamefont{{Dav{\' e}}}},
  \bibinfo{author}{\bibfnamefont{D.~N.} \bibnamefont{{Spergel}}},
  \bibinfo{author}{\bibfnamefont{P.~J.} \bibnamefont{{Steinhardt}}},
  \bibnamefont{and} \bibinfo{author}{\bibfnamefont{B.~D.}
  \bibnamefont{{Wandelt}}}, \bibinfo{journal}{\apj}
  \textbf{\bibinfo{volume}{547}}, \bibinfo{pages}{574} (\bibinfo{year}{2001}).

\bibitem[{\citenamefont{{Colladay} and {{Kosteleck\'{y}}, V. A.}}(1998)}]{ck98}
\bibinfo{author}{\bibfnamefont{D.}~\bibnamefont{{Colladay}}} \bibnamefont{and}
  \bibinfo{author}{\bibnamefont{{{Kosteleck\'{y}}, V. A.}}},
  \bibinfo{journal}{Phys. Rev. D} \textbf{\bibinfo{volume}{58}},
  \bibinfo{pages}{116002} (\bibinfo{year}{1998}).

\bibitem[{\citenamefont{{Bertolami} and {Carvalho}}(2000)}]{bc00}
\bibinfo{author}{\bibfnamefont{O.}~\bibnamefont{{Bertolami}}} \bibnamefont{and}
  \bibinfo{author}{\bibfnamefont{C.~S.} \bibnamefont{{Carvalho}}},
  \bibinfo{journal}{\prd} \textbf{\bibinfo{volume}{61}},
  \bibinfo{pages}{103002} (\bibinfo{year}{2000}).

\bibitem[{\citenamefont{{Bertolami}}(2000)}]{o00}
\bibinfo{author}{\bibfnamefont{O.}~\bibnamefont{{Bertolami}}},
  \bibinfo{journal}{Nucl. Phys. B-Proc. Sup.} \textbf{\bibinfo{volume}{88}},
  \bibinfo{pages}{49} (\bibinfo{year}{2000}).

\bibitem[{\citenamefont{{Alfaro} et~al.}(2002)\citenamefont{{Alfaro},
  {Morales-T{\' e}cotl}, and {Urrutia}}}]{amu02}
\bibinfo{author}{\bibfnamefont{J.}~\bibnamefont{{Alfaro}}},
  \bibinfo{author}{\bibfnamefont{H.~A.} \bibnamefont{{Morales-T{\' e}cotl}}},
  \bibnamefont{and} \bibinfo{author}{\bibfnamefont{L.~F.}
  \bibnamefont{{Urrutia}}}, \bibinfo{journal}{\prd}
  \textbf{\bibinfo{volume}{65}}, \bibinfo{pages}{103509}
  (\bibinfo{year}{2002}).

\bibitem[{\citenamefont{{Bastero-Gil} and {Mersini}}(2003)}]{bm03}
\bibinfo{author}{\bibfnamefont{M.}~\bibnamefont{{Bastero-Gil}}}
  \bibnamefont{and}
  \bibinfo{author}{\bibfnamefont{L.}~\bibnamefont{{Mersini}}},
  \bibinfo{journal}{\prd} \textbf{\bibinfo{volume}{67}},
  \bibinfo{pages}{103519} (\bibinfo{year}{2003}).

\bibitem[{\citenamefont{{Alexander} and {Magueijo}}(2001)}]{sm01}
\bibinfo{author}{\bibfnamefont{S.~H.} \bibnamefont{{Alexander}}}
  \bibnamefont{and}
  \bibinfo{author}{\bibfnamefont{J.}~\bibnamefont{{Magueijo}}}
  (\bibinfo{year}{2001}), \eprint{hep-th/0104093}.

\bibitem[{\citenamefont{{Kosteleck\'{y}} and
  {Samuel}}(1989{\natexlab{a}})}]{ks89a}
\bibinfo{author}{\bibfnamefont{V.~A.} \bibnamefont{{Kosteleck\'{y}}}}
  \bibnamefont{and} \bibinfo{author}{\bibfnamefont{S.}~\bibnamefont{{Samuel}}},
  \bibinfo{journal}{Phys. Rev. D} \textbf{\bibinfo{volume}{40}},
  \bibinfo{pages}{1886} (\bibinfo{year}{1989}{\natexlab{a}}).

\bibitem[{\citenamefont{{Kosteleck\'{y}} and
  {Samuel}}(1989{\natexlab{b}})}]{ks89b}
\bibinfo{author}{\bibfnamefont{V.~A.} \bibnamefont{{Kosteleck\'{y}}}}
  \bibnamefont{and} \bibinfo{author}{\bibfnamefont{S.}~\bibnamefont{{Samuel}}},
  \bibinfo{journal}{Phys. Rev. D} \textbf{\bibinfo{volume}{39}},
  \bibinfo{pages}{683} (\bibinfo{year}{1989}{\natexlab{b}}).

\bibitem[{\citenamefont{{Bertolami}}(1997)}]{b97}
\bibinfo{author}{\bibfnamefont{O.}~\bibnamefont{{Bertolami}}},
  \bibinfo{journal}{Class. Quant. Grav.} \textbf{\bibinfo{volume}{14}},
  \bibinfo{pages}{2785} (\bibinfo{year}{1997}).

\bibitem[{\citenamefont{{Frey}}(2003)}]{f03}
\bibinfo{author}{\bibfnamefont{A.~R.} \bibnamefont{{Frey}}},
  \bibinfo{journal}{JHEP} \textbf{\bibinfo{volume}{4}}, \bibinfo{pages}{12}
  (\bibinfo{year}{2003}).

\end{thebibliography}
\end{document}